# Wide-Area GNSS Interference Monitoring with CYGNSS GNSS-R Delay-Doppler Noise Floor Observations

Ji-Hyeon Shin and Pyo-Woong Son, *Member, IEEE*

***Abstract*— Global Navigation Satellite System (GNSS) interference increasingly threatens positioning, navigation, and timing services and requires monitoring over areas beyond dense ground networks. Low Earth orbit satellites provide complementary regional coverage, and previous studies have shown that terrestrial radio-frequency interference (RFI) produces measurable anomalies in spaceborne GNSS and GNSS-reflectometry observations. The Cyclone Global Navigation Satellite System (CYGNSS) has been used to map and characterize interference, but existing approaches generally depend on long-term data accumulation, full Delay-Doppler Map (DDM) processing, or special raw intermediate-frequency acquisitions and lack epoch-level validation against independent data. This paper presents a lightweight detector using four channel-wise DDM noise-floor values routinely distributed in CYGNSS Level-1 products. Because the channels observe different reflected signals through two nadir antennas, interference-related elevations may be strongly asymmetric. The detector uses the channel maximum to preserve these responses and applies temporal-persistence or multi-satellite-concurrence screening to reject isolated anomalies. Evaluation at the White Sands Missile Range used an independent reference formed from Federal Aviation Administration Notices to Air Missions and concurrent Automatic Dependent Surveillance–Broadcast navigation-integrity degradation. Relative to the single-epoch mean baseline, the proposed detector increased the probability of detection by 0.063 at the same observed false-alarm rate, with a 95% confidence interval of [0.034, 0.094]. In the Middle East, it also detected both vertical-stripe and atypical elevated-background DDM structures. These results demonstrate lightweight epoch-level monitoring of wide-area or temporally sustained GNSS interference using routine Level-1 products.**

This work was supported in part by the Korea Agency for Infrastructure Technology Advancement (KAIA) grant funded by the Ministry of Land, Infrastructure and Transport (Grant No. RS-2026-25527966), and in part by the Regional Innovation System & Education (RISE) program through the Chungbuk Regional Innovation System & Education Center, funded by the Ministry of Education (MOE) and Chungcheongbuk-do, Republic of Korea (Grant No. 2025-RISE-11-014-03). (Corresponding author: Pyo-Woong Son.)

Ji-Hyeon Shin is with the Department of Electronic Engineering, Chungbuk National University, Cheongju 28644, South Korea (e-mail: sjh3251@cbnu.ac.kr).

Pyo-Woong Son is with the Department of Electronic Engineering, Chungbuk National University, Cheongju 28644, South Korea (e-mail: pwson@cbnu.ac.kr).

## I. INTRODUCTION

Global Navigation Satellite Systems (GNSS) have become critical infrastructure for positioning, navigation, and timing (PNT) across civil and military domains, including transportation, financial systems, telecommunications, geodetic surveying, and autonomous platforms [1]–[5]. Because GNSS signals are received at extremely low power levels, they are inherently vulnerable to radio frequency interference (RFI), and even modest interference can degrade receiver tracking or deny service altogether [4]–[6]. Recent reports indicate that GNSS jamming and spoofing are no longer isolated anomalies but form part of a broader and increasingly persistent disruption pattern, with sustained activity documented in several geopolitically sensitive regions [7]–[10]. Complementary terrestrial systems such as enhanced Loran (eLoran) and medium-frequency R-Mode can maintain PNT service during GNSS disruption [11]–[15], but they do not themselves reveal where or when interference occurs. Detecting and characterizing interference is therefore a separate requirement that must be addressed over large geographic regions.

Ground-based interference-monitoring networks can detect and localize RFI sources, but their deployment and maintenance are costly, and their coverage is limited to instrumented areas [16]–[19]. Low Earth orbit (LEO) satellites can observe GNSS-related signals over regions substantially larger than those covered by dense ground networks [8]–[10]. The Cyclone Global Navigation Satellite System (CYGNSS) constellation is well suited to this application. Its eight observatories provide repeated coverage within the principal low-latitude band, and each carries a Delay Doppler Mapping Instrument (DDMI) with one zenith and two nadir-pointing antennas. The instrument routinely generates Level-1 Delay-Doppler Map (DDM) products from four simultaneous reflected-signal channels [20]–[22].

Terrestrial interference leaves measurable signatures in spaceborne GNSS observations. Onboard receivers have detected and characterized interference using raw intermediate-frequency data collected on the International Space Station and signal-to-noise-ratio variations recorded by Blackjack/TriG receivers on radio-occultation missions [8]–[10]. GNSS reflectometry (GNSS-R), which processes reflected GNSS signals into DDMs for the retrieval of ocean wind, land-surface, and soil-moisture information [21], [23]–[25], is sensitive to the same interference. TechDemoSat-1 observations showed anomalous noise behavior attributable to terrestrial sources [26], while CYGNSS signal-characterization work demonstrated that the DDM background contains a variable component responsive to external L-band radiation in addition to relatively stable thermal contributions [27]. These findings establish that terrestrial interference can produce measurable abnormalities in the background-noise structure of spaceborne GNSS-R observations.

Within CYGNSS specifically, a multi-year study aggregated Level-1 observations collected from 2017 to 2022 to produce global maps of interference hotspots and estimate source locations and transmission durations [7]. Subsequent studies characterized and geolocated interference sources using raw intermediate-frequency data and Level-1 products [28]–[30], while other work quantified RFI-induced degradation of GNSS-R signal quality and investigated mitigation at the receiver or product level [31]–[33]. These approaches provide detailed spatial, morphological, or signal-level information, but they rely on long-term data accumulation, special raw-IF acquisitions, full DDM data, or additional source-estimation processing. They therefore do not directly address lightweight detection that can generate a decision at each routinely delivered Level-1 observation epoch with minimal computation.

Single-epoch detection introduces a distinct aggregation problem. Long-term mapping accumulates a large number of observations, so observation-specific variations—including delay-reference mismatch, terrain-dependent background elevation, instrument-state variation, and geometry-dependent gain differences [27], [38], [41]—are diluted across the accumulated dataset. At a single epoch, this averaging across observations is unavailable, and the four simultaneous channel values must instead be reduced to one decision variable. These channels are not redundant measurements of the same reflected signal under identical conditions; they are derived from different reflected GNSS signals received through two differently oriented nadir antennas, with their selection determined by the instantaneous observation geometry [20]–[22]. A ground-based interference source may consequently elevate only a subset of the channels. Reliable single-epoch detection therefore requires an aggregation rule that preserves interference-related elevations expressed in only part of the four-channel observation while avoiding excessive sensitivity to isolated non-interference anomalies.

An additional gap concerns quantitative detector validation. Existing CYGNSS studies have not reported epoch-level probability of detection and false-alarm rate against a reference obtained independently of CYGNSS. Their results have primarily been presented as interference maps, candidate source locations, estimated transmission durations, or source characterization from special raw-IF acquisition modes [7], [28], [29]. Without an independent reference, an increase in detector-positive observations cannot be distinguished from an increase in false alarms, and competing detector designs cannot be evaluated on a common quantitative basis.

To address these gaps, this study develops a lightweight epoch-level detector using only the four channel-wise noise-floor values routinely distributed in CYGNSS Level-1 products. The detector uses the channel maximum to preserve interference-related elevations that appear asymmetrically across the simultaneous channels and applies additional screening to suppress isolated anomalies that are inconsistent with spatially extensive or temporally sustained interference. The single-epoch mean of the same four measurements is retained as a baseline, and detector performance is evaluated against an independent aviation-domain reference constructed from Federal Aviation Administration Notices to Air Missions (NOTAMs) and Automatic Dependent Surveillance–Broadcast (ADS-B) navigation-integrity reports.

The main contributions of this work are as follows:

1. Physical interpretation and empirical characterization of the channel-wise asymmetry that frequently accompanies interference-related noise-floor elevation in the four simultaneous CYGNSS observations.
2. A lightweight maximum-based Level-1 detector that improves sensitivity to partial-channel interference while using additional screening to suppress isolated anomalies.
3. An epoch-level cross-domain validation framework that combines spaceborne CYGNSS observations with independent NOTAM and ADS-B records to quantify detector performance using probability of detection, false-alarm rate, and paired statistical analysis.

The remainder of this paper is organized as follows. Section II describes the CYGNSS observation structure and establishes the DDM noise floor as an interference-sensitive, but not interference-exclusive, observable. Section III presents the proposed epoch-level detection and screening framework. Section IV describes the experimental design, distribution-based threshold determination, and evaluation results. Section V discusses the physical interpretation, practical implications, and limitations of the findings. Section VI concludes the paper.

## II. CYGNSS Observation Structure and DDM Noise Floor as an Interference-Sensitive Observable

### *A. CYGNSS Observation Structure and DDM Formation*

The CYGNSS constellation consists of eight low Earth orbit (LEO) observatories at an altitude of approximately 510–520 km and an inclination of 35°, providing repeated coverage within the principal low-latitude band. Each observatory carries a DDMI comprising one zenith antenna for direct-signal reception and two nadir-pointing antennas for Earth-surface-reflected GNSS signals. At each observation epoch, the receiver considers multiple candidate reflections and routinely forms four simultaneous reflected-signal channels for Level-1 DDM generation [38]. This four-channel observation structure is central to the present study because it defines the measurement geometry from which the DDM noise-floor observable is derived.

A DDM represents the bistatic correlation power of a reflected GNSS signal on a two-dimensional plane of code delay and Doppler frequency [39]. Both axes are referenced to the specular point, i.e., the surface location corresponding to the minimum propagation path between the GNSS transmitter and the CYGNSS receiver. Under nominal conditions, the strongest reflected power is concentrated near the specular region, while power from surrounding scattering areas is distributed over neighboring delay-Doppler bins according to excess path length and relative motion. Fig. 1 illustrates the reflection geometry and the definition of the specular reference used in CYGNSS DDM processing.

Because the four simultaneous channels are derived from different reflected GNSS signals and are observed through nadir antennas, they are not redundant measurements of the same reflection under identical reception conditions [20]–[22]. Instead, they represent four concurrent but geometrically distinct reflected-signal observations. These channel-to-channel geometric differences are the basis of the aggregation problem addressed in Section III.

### *B. DDM Noise Floor as an Interference-Sensitive Observable*

A Delay-Doppler Map (DDM) represents the bistatic correlation power of a reflected GNSS signal over code delay and Doppler frequency [39]. Both axes are referenced to the specular point, which corresponds to the surface location having the minimum propagation path between the GNSS transmitter and the CYGNSS receiver. Under nominal conditions, the strongest reflected power is concentrated near the specular region, while contributions from surrounding scattering areas are distributed over neighboring delay-Doppler bins according to their excess path lengths and relative motion [23], [40]. A representative nominal DDM is shown in Fig. 2(a), where the dominant reflected-signal power remains confined to the expected scattering region around the specular point.

Because the specular point defines the minimum-delay propagation path, nominal reflected-signal power should not appear at delays earlier than the specular delay when the geometric reference is correctly established. The negative-delay region is therefore commonly treated as a forbidden zone, in which primarily background-level contributions are expected under nominal conditions [27]. In the CYGNSS Level-1 product, the DDM noise floor is calculated as the average signal level over these forbidden-zone bins and is reported in instrument count units rather than as an absolute calibrated power quantity [38], [41].

External radio-frequency interference can increase the received background power within the DDM, including the forbidden zone. Previous CYGNSS signal-characterization studies have shown that the DDM background contains a variable component responsive to external L-band radiation in addition to relatively stable thermal contributions [27]. Interference may therefore appear as an elevated forbidden-zone noise floor. As illustrated by the abnormal examples in Fig. 2(b), such contamination can produce a pronounced enhancement at a particular Doppler frequency, resulting in the vertical-stripe structure commonly reported in GNSS-R RFI studies [7], [31]–[33].

However, RFI does not produce a unique DDM morphology. As shown in Fig. 2(b), contaminated DDMs may exhibit diffuse or irregular background elevation without a clearly defined Doppler-localized structure, in addition to the canonical vertical-stripe pattern. A detector based only on a specific image pattern may consequently fail to identify interference that appears in a different form. The forbidden-zone average provides a more general interference-sensitive observable because both canonical and atypical contamination can increase power outside the nominal scattering region.

Conversely, an elevated forbidden-zone noise floor does not by itself establish the presence of RFI. The DDM delay reference is determined from assumed reflection geometry and surface-elevation information, and mismatch between the assumed and actual geometry can shift nominal reflected power into the negative-delay region. Topographic complexity, scene-dependent scattering, instrument-state variation, and other observation-specific effects may also elevate the estimated background in the absence of interference. These effects are generally more significant over land than over open ocean.

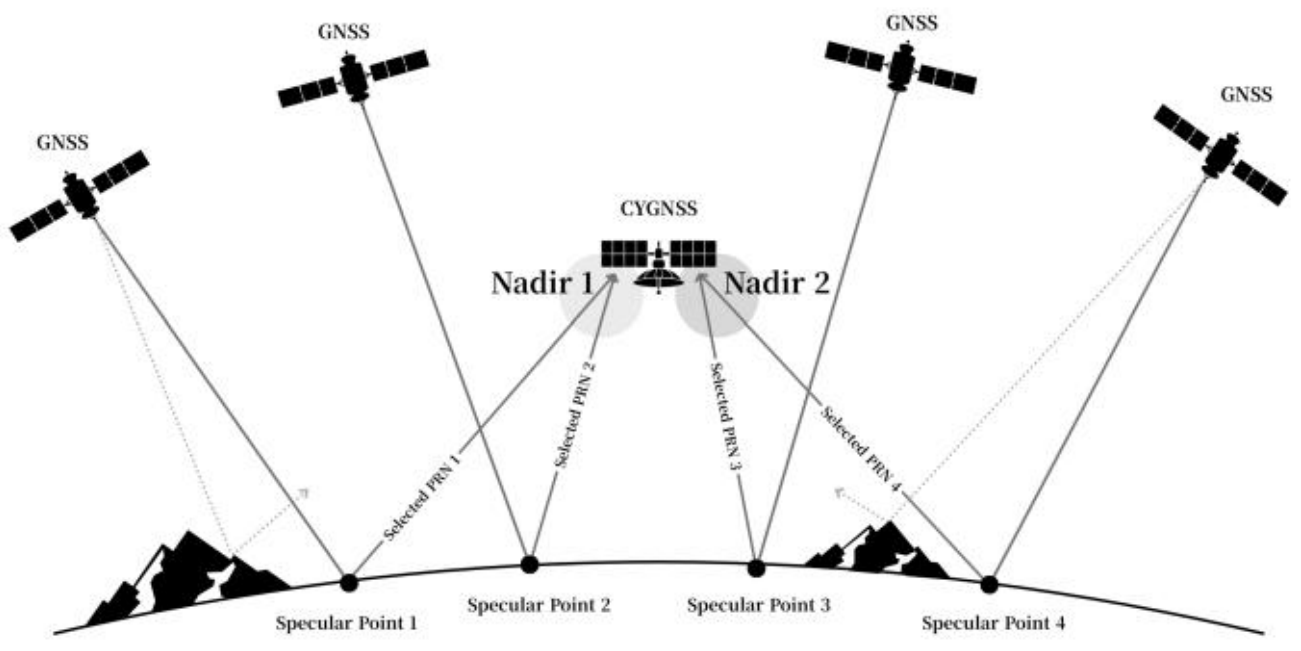

**Fig. 1.** Geometry of GNSS signal reflection and definition of the specular point, which provides the minimum propagation path between the GNSS transmitter and the CYGNSS receiver.

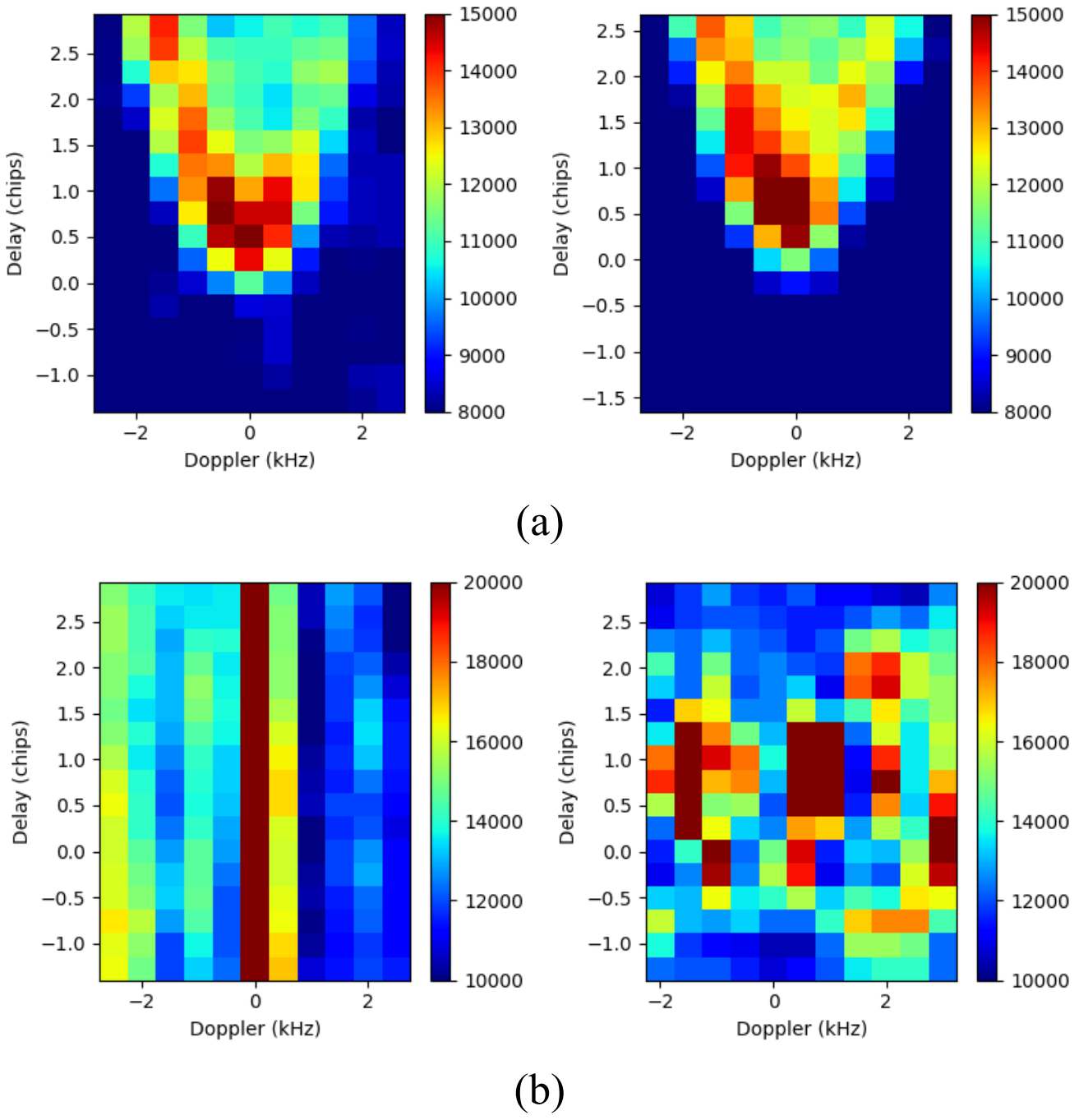

**Fig. 2.** Representative CYGNSS DDM patterns: (a) nominal reflected-signal DDM, in which the strongest power is concentrated near the specular region, and (b) abnormal DDMs exhibiting interference-related contamination, including typical vertical-stripe and atypical elevated-background patterns.

The forbidden-zone noise floor should therefore be interpreted as an interference-sensitive, but not interference-exclusive, observable. Reliable epoch-level monitoring requires not only retaining genuine interference-related elevations but also rejecting isolated or scene-induced anomalies that produce similar noise-floor increases. The detection framework in Section III addresses this distinction by combining the channel-wise noise-floor response with temporal-persistence and multi-satellite-concurrence screening.

## III. Proposed Detection Framework

### *A. Maximum-Based Aggregation Under Channel-Asymmetric Responses*

DDM abnormalities can be identified in several ways, including image-based examination of the full DDM, signal-level analysis using raw intermediate-frequency data, and statistical analysis of observations accumulated over long time periods [7], [28], [29], [31]–[36]. These approaches can provide detailed information on interference morphology, signal characteristics, or persistent spatial patterns. However, they generally require full DDM data, special raw-IF acquisitions, image-level processing, or the accumulation of a large number of observations. They are therefore less suited to immediate and routine detection at each observation epoch. For continuous wide-area monitoring, a computationally lightweight method is needed that can operate directly on routinely distributed Level-1 measurements and produce a decision from a single epoch.

The CYGNSS Level-1 product provides four simultaneous forbidden-zone noise-floor values at each epoch. These channels are derived from different reflected GNSS signals and are received through nadir antennas, so they do not represent redundant measurements under identical observation conditions [20]–[22]. Differences in reflection geometry, antenna orientation, and source visibility can cause the four channels to respond unequally to the same ground-based interference source. Consequently, a source may produce a clear noise-floor elevation in only one or two channels while leaving the remaining channels comparatively unaffected.

Throughout this paper, the Level-1 forbidden-zone noise floor, originally reported in instrument counts, is expressed on a logarithmic scale as $10\log_{10}$ of the count value. Let $x_{s,p}(t)$ denote the noise-floor value in dB for CYGNSS observatory $s$, channel $p \in \{1,2,3,4\}$, at epoch $t$. Averaging the four values observed at a single epoch gives the mean statistic,

$$\bar{n}_s(t) = \frac{1}{4}\sum_{p=1}^{4} n_{s,p}(t) \tag{1}$$

This statistic is stable when the four channels respond similarly, but it attenuates the response of a minority of strongly contaminated channels when the interference exposure is uneven. Under such conditions, the resulting statistic may fall below threshold even when one or more channels contain a clear interference signature.

Fig. 3 presents a representative observed case that illustrates this effect numerically. In this example, the four simultaneous channel noise-floor values are 36, 38, 39, and 43 dB, respectively, and the adopted threshold is 41 dB. Although one channel clearly exceeds threshold, the four-channel mean is reduced to 39 dB by the three less-affected channels and therefore fails to trigger detection. By contrast, the four-channel maximum remains 43 dB and preserves the exceedance.

To avoid this structural sensitivity loss, the detector uses the largest of the four simultaneous channel values,

$$m_s(t) = \max_{p\in\{1,2,3,4\}} n_{s,p}(t) \tag{2}$$

The maximum uses the same Level-1 noise-floor measurements [27], [38], [41] while preserving the strongest channel elevation. It addresses sensitivity loss under channel-asymmetric elevations only; whether a candidate is consistent with a wide-area or temporally sustained event is determined by the screening conditions in Section III-C.

### *B. Threshold-Based Candidate Flagging*

Candidate RFI events are identified by thresholding this value. A candidate flag for satellite s at epoch t is defined as

$$c_s(t) = \begin{cases} 1, m_s(t) > \tau \\ 0, m_s(t) \le \tau \end{cases} \tag{3}$$

where τ denotes the DDM-noise-floor threshold expressed in dB. In the present work, τ=41 dB is adopted as the operating threshold. Section IV-A describes how this value was established experimentally from reference regions of comparatively low interference activity, and Section IV-C reports its subsequent evaluation against the WSMR reference standard.

For comparison, the baseline flag obtained from the single-epoch mean of the same four values is defined as

$$\bar{c}_s(t) = \begin{cases} 1, \bar{n}_s(t) > \tau \\ 0, \bar{n}_s(t) \le \tau \end{cases} \tag{4}$$

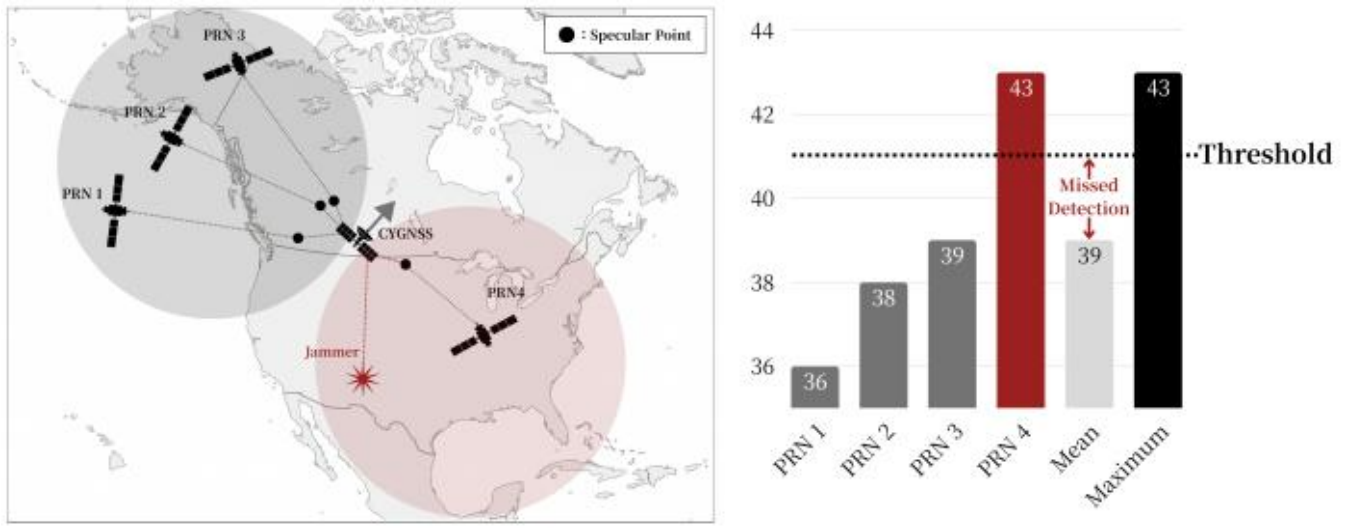

**Fig. 3.** Conceptual illustration of the improved sensitivity of maximum-based aggregation to partial-channel interference.

Because (3) and (4) use the same measurement and the same threshold, any difference between their outputs comes from the choice of decision variable alone.

The detector requires four scalar noise-floor values, one comparison with the threshold, and the subsequent logical screening. Its computational cost is therefore fixed per epoch, and no special raw intermediate-frequency acquisition [29] or additional source-estimation processing [28] is needed.

### *C. Screening for Wide-Area or Persistent Events*

The maximum preserves sensitivity to channel-asymmetric interference, but by construction it is also more responsive to isolated single-channel excursions than an averaging statistic. Such excursions may arise from transient receiver artifacts, local geometric mismatch, or other non-RFI effects that perturb the DDM background [7], [32], [33], [38], [41]. Because the detector is intended to identify high-intensity interference that is spatially extensive or temporally sustained, rather than to flag every anomalous epoch, candidate exceedances identified by $c_s(t)$are subjected to an additional screening stage before being retained as final detections.

Two complementary screening conditions are applied. The first is multi-satellite concurrence. If two or more CYGNSS observatories independently record candidate exceedances at the same epoch within a given analysis case, the event is retained as a wide-area interference candidate supported by multiple observatories. This condition is motivated by the expectation that sufficiently strong or spatially extensive ground-based interference can be visible to multiple satellites whose observation footprints lie within the affected region.

The second condition is temporal persistence. If a candidate exceedance is observed by only one satellite, it is retained only when the exceedance persists continuously over a 10 s interval along the satellite track. Defining the persistence indicator for satellite s as

$$p_s(t) = \begin{cases} 1, c_s(t') = 1 \text{ for all } t' \in [t - 10 \text{ s},\, t] \\ 0, \mathit{otherwise} \end{cases} \tag{5}$$

the 10 s requirement serves as a conservative minimum-duration condition for sustained events. At CYGNSS altitude, the change in slant range over such an interval is small relative to the full line-of-sight distance, so a ground-based source that is detectable from orbit would be expected to remain observable for substantially longer than this interval. The persistence criterion therefore functions as a practical filter against short-lived isolated anomalies without imposing a restrictive event model.

To express the concurrence condition, let

$$q(t) = \begin{cases} 1, \sum_s c_s(t) \ \geq 2 \\ 0, \sum_s c_s(t) < 2 \end{cases} \tag{6}$$

Then the final retained detection for satellite s is defined as

$$d_s(t) = \begin{cases} 1, c_s(t) = 1 \text{ and } [q(t) = 1 \text{ or } p_s(t) = 1] \\ 0, \text{otherwise} \end{cases} \tag{7}$$

Final detections must therefore satisfy either multi-satellite concurrence or 10 s persistence; isolated single-epoch exceedances are rejected. Section IV evaluates the detector in a documented RFI environment and in a persistent-interference environment, with the operational kurtosis flag and the mean baseline as points of comparison.

## IV. Experimental Results

### *A. Experimental Design and Threshold Determination*

The detector was evaluated using NASA CYGNSS Level-1 Science Data Record Version 3.2 over 1–24 May 2025 in two complementary regions. The White Sands Missile Range (WSMR; 26.5°–39°N, 244°–264°E) provides a documented setting, with GPS RFI tests conducted during periods announced through Federal Aviation Administration Notices to Air Missions (NOTAMs) [42]–[46], and it does not appear as a persistently elevated background area in the one-year CYGNSS noise-floor climatology of Fig. 4 (July 2025–June 2026). The Middle East (29°–37°N, 34°–60°E) was selected as a persistent-interference environment: frequent GNSS interference has been reported there in previous spaceborne monitoring studies [7], [10], and it does appear as a persistently elevated region in the same climatology. The Pacific Ocean and Amazon regions used below for threshold characterization lie in the lower part of that global distribution.

The operating threshold was determined from two reference regions where the DDM noise floor remains comparatively low and no persistent RFI has been reported: the Pacific Ocean (30°–35°N, 184°–186°E) and the Amazon rainforest (2°–6°S, 280°–286°E), both analyzed from July 2025 to June 2026. The Pacific gives an open-ocean case; the Amazon gives a land case in which non-interference noise-floor elevation may still occur through topographic complexity and delay-reference mismatch. Fig. 5 shows the two distributions and Table I the corresponding sample counts and exceedance statistics. These regions serve as low-interference references for a detector targeting sustained, high-intensity RFI.

In the Pacific dataset none of the 1,342,201 samples exceeded 41 dB; in the Amazon dataset 2,699 of 1,095,538 samples (0.2%) did. Raising the threshold from 40 to 41 dB reduced exceedances across the two reference datasets by about 55%. We therefore adopted 41 dB as the common threshold for both the proposed detector and the mean baseline, and its sensitivity was subsequently evaluated against the WSMR reference standard in Section IV-C.

Three configurations were compared under identical data and regional conditions. The proposed detector and the mean baseline share the 41 dB threshold and differ only in the decision variable; the kurtosis flag distributed with the CYGNSS Level-1 product [7], [37] uses its own fixed criterion, kurtosis > 4.0.

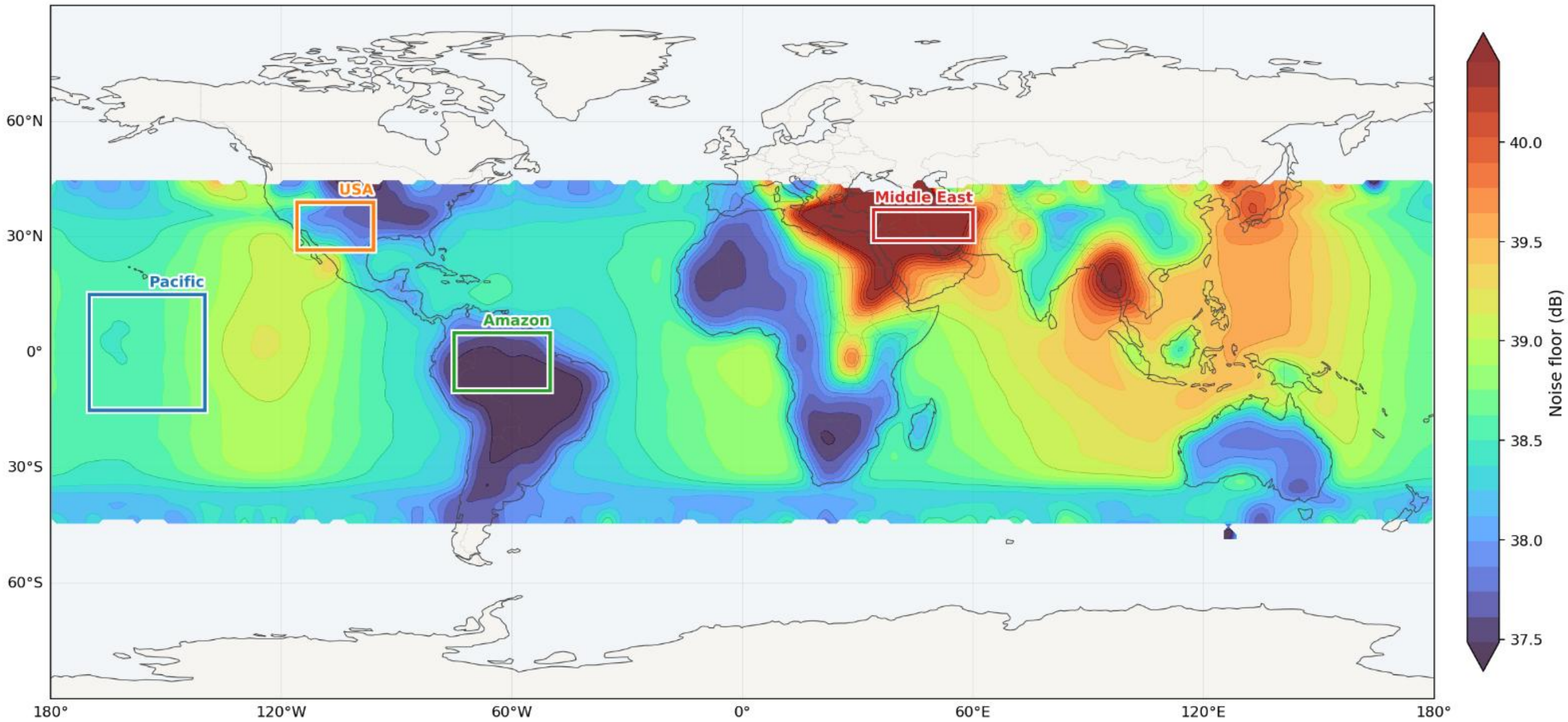

**Fig. 4.** CYGNSS GNSS-R DDM noise floor from July 2025 to June 2026, showing elevated background levels over interference-prone regions such as the Middle East, and comparatively low background levels over WSMR and over the two low-interference reference regions used for threshold setting.

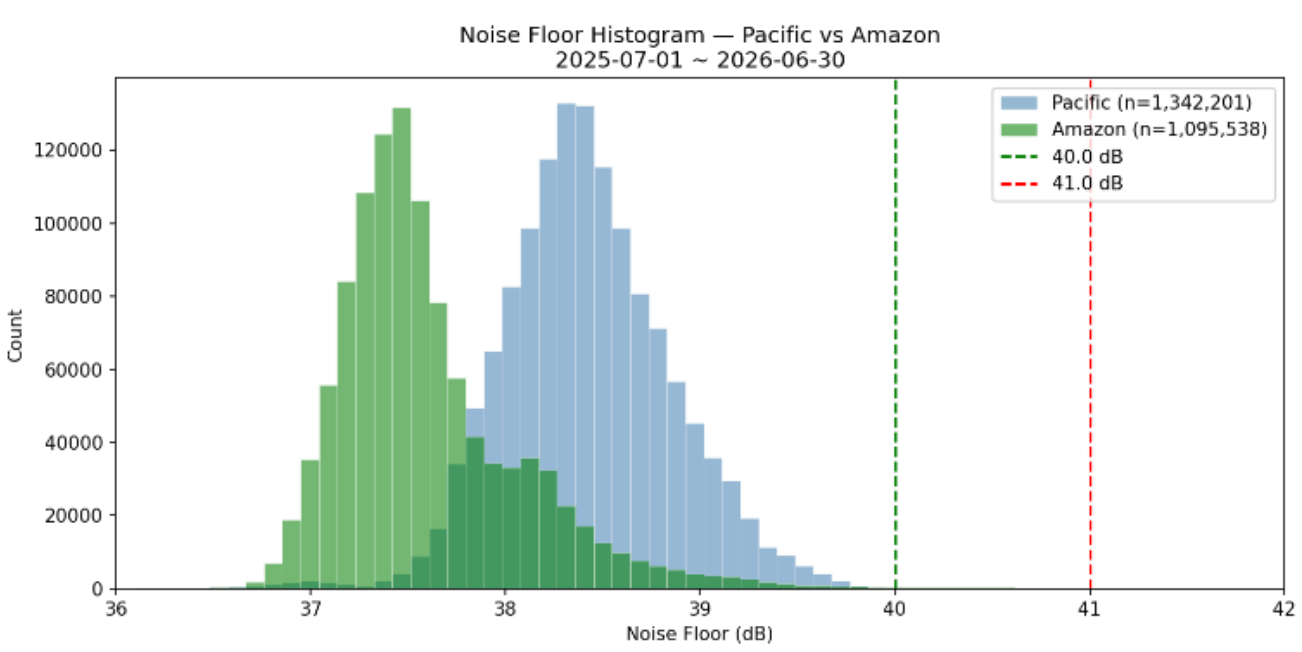

**Fig. 5.** DDM noise-floor distributions in the two low-interference reference regions used for threshold setting: the Pacific Ocean and the Amazon rainforest. The dashed lines indicate the 40 dB and 41 dB thresholds used to compare threshold-exceedance behavior in the two reference regions.

TABLE I
SUMMARY STATISTICS OF REFERENCE-ENVIRONMENT NOISE-FLOOR DISTRIBUTIONS USED FOR THRESHOLD SETTING

| Region | Total Samples | 40 < NF ≤ 41 dB | Samples >41 dB |
|---|---|---|---|
| Pacific | 1,342,201 | 146 | 0 |
| Amazon | 1,095,538 | 3,116 | 2,699 |

*B. Sensitivity to Channel-Asymmetric Interference Under Documented RFI Conditions*

The WSMR case provides a documented interference-active setting in which detector behavior can be examined against announced jamming periods [42]–[46].

Fig. 6 compares the daily numbers of flagged events over the WSMR region from 1 to 24 May 2025 for the kurtosis flag, the mean baseline, and the proposed detector. A clear difference is observed in how closely the three configurations align with the documented NOTAM windows. The two noise-floor-based configurations concentrate their detections primarily within the announced RFI periods, whereas the kurtosis flag produces high counts on essentially every date, with little relation to the NOTAM schedule. This contrast is consistent with previous findings that kurtosis is sensitive to abnormal DDM structure but not always selective for actual interference [7]. Notably, on 14, 15, and 20 May, the proposed detector produced clear threshold exceedances where the mean baseline produced few or none.

Fig. 7 presents representative DDMs from those three dates that were retained by the proposed detector but not by the mean baseline. All three occurred within the WSMR region during announced RFI windows and show broad power elevation in the forbidden zone, where nominal reflected-signal power is not expected under correct geometric referencing, rather than isolated single-bin fluctuations. Against the otherwise low WSMR background, such elevations during announced RFI windows are difficult to explain by nominal reflected-signal behavior alone.

Among the WSMR epochs retained by the proposed detector but missed by the mean baseline, the median channel values, ordered from the minimum to the maximum within each epoch, were 39.03, 39.51, 40.50, and 41.63 dB. The 2.60 dB difference between the median minimum and maximum corresponds to a factor of approximately 1.82 in instrument counts, indicating that the strongest-channel response was about 82% higher than the weakest-channel response. The median four-channel mean for the same epochs was 40.24 dB, below the 41 dB threshold, whereas the median maximum remained above it. These statistics directly show that substantial channel asymmetry can cause an interference-related elevation to be diluted below threshold by mean-based aggregation. The antenna-group-associated character of these responses is also evident at the population level. Across the WSMR observations, antenna-group asymmetry—defined as a difference of at least 1 dB between the mean noise floors of the two nadir-antenna groups—occurred in 14.9% of epochs below 40 dB and in 75.5% of epochs at or above 40 dB. The epochs additionally retained by the proposed detector therefore coincided with documented RFI periods and exhibited channel-asymmetric background elevations that were susceptible to dilution by averaging.

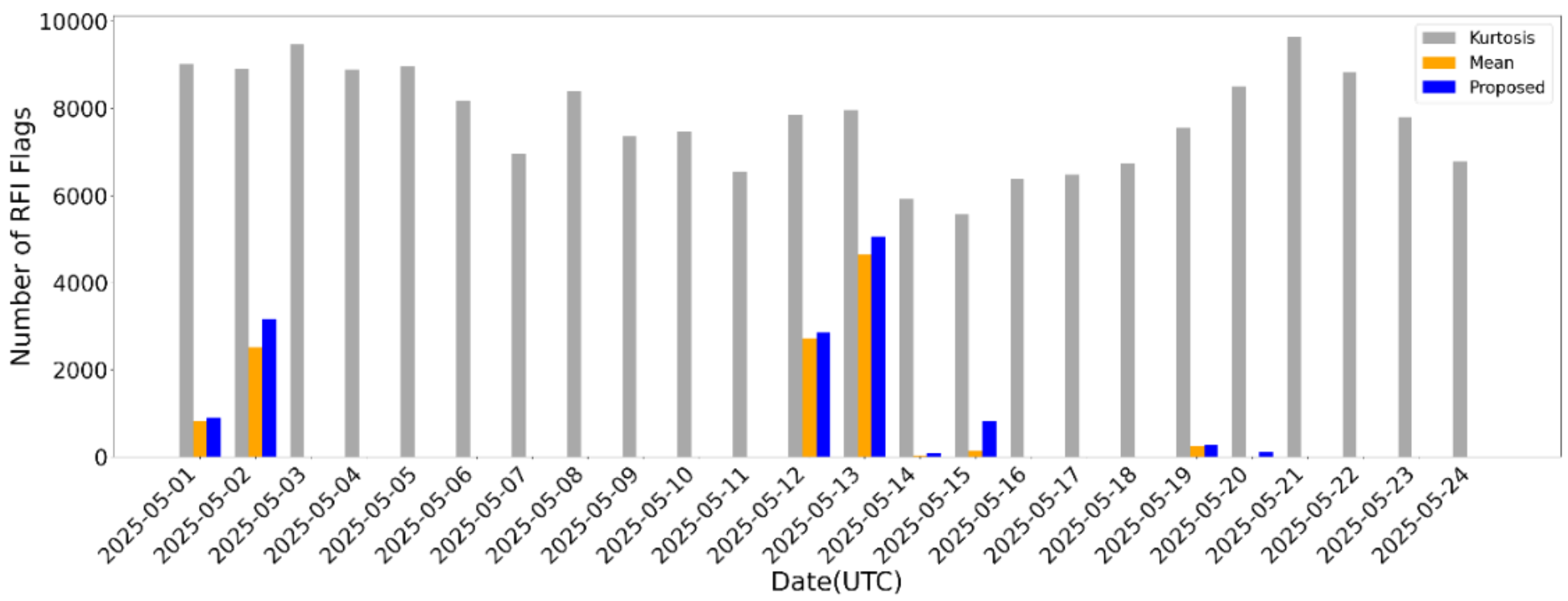

**Fig. 6.** Daily RFI flag counts over the White Sands Missile Range from May 1 to 24, 2025, for the kurtosis flag, the mean baseline, and the proposed detector.

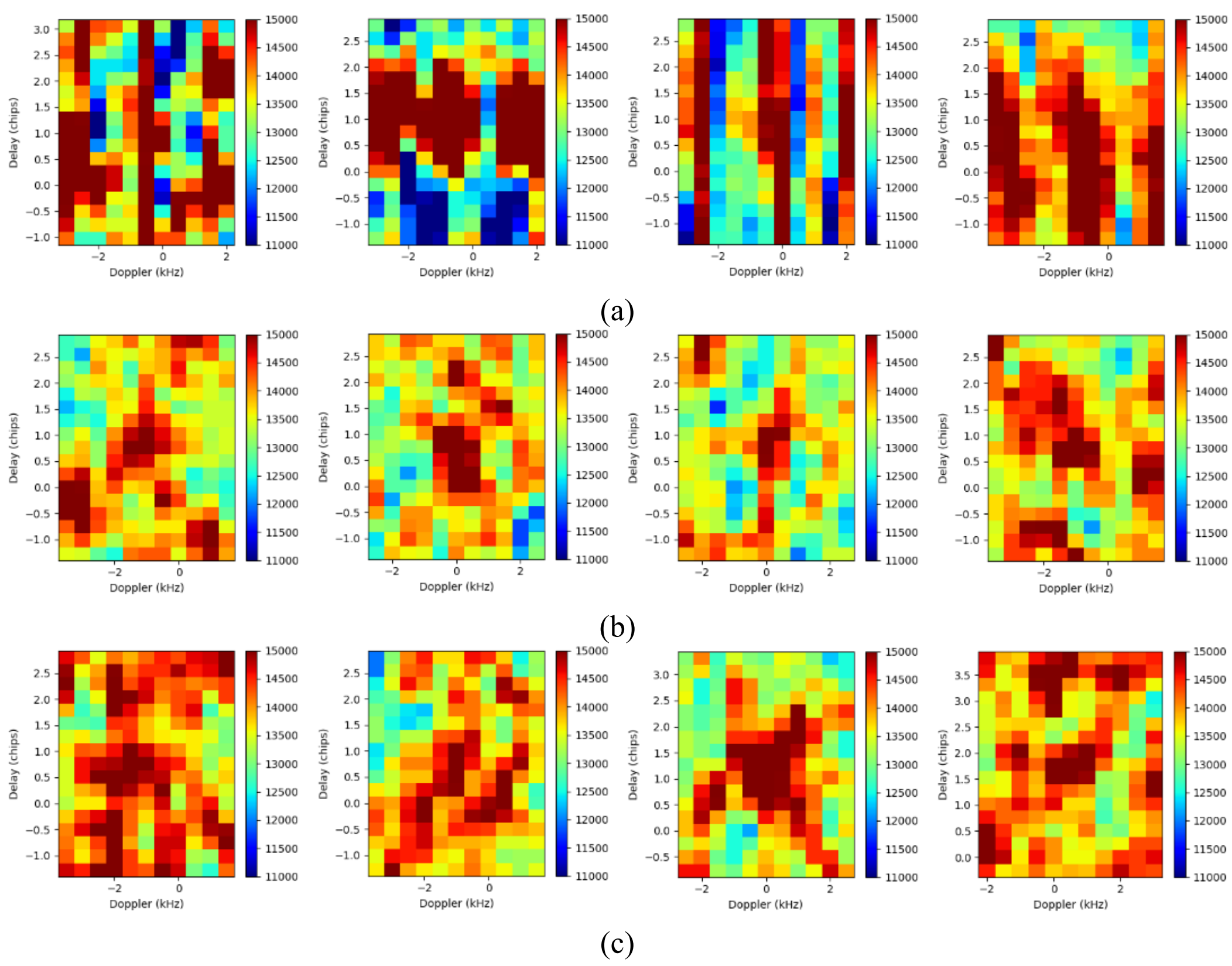

**Fig. 7.** Representative abnormal DDMs from May 14 (a), 15 (b), and 20 (c) at WSMR that were retained by the proposed detector but not by the mean baseline.

### *C. Detection Performance at WSMR Against NOTAM and ADS-B Records*

An increase in the number of flagged epochs alone cannot distinguish improved interference detection from more frequent false alarms. The WSMR results are therefore evaluated against a reference standard that pairs the FAA Notice to Air Missions (NOTAM) schedule with Automatic Dependent Surveillance–Broadcast (ADS-B) reports obtained from the OpenSky Network, so that an announced test window is treated as interference-active only when aircraft in the region concurrently report degraded navigation integrity. Both data sources are aviation-domain records acquired independently of CYGNSS. Because CYGNSS observations are reported at 0.5 s intervals and the ADS-B condition is evaluated at 1 s resolution, a second is counted as a CYGNSS detection if either of the two 0.5 s samples within that second is flagged; all results in this subsection are therefore reported as 1 s binary evaluation outcomes.

Individual aircraft reports are classified using the Navigation Integrity Category (NIC), an ADS-B integrity indicator related to the containment radius of the reported position. Section 91.227 of Title 14 of the Code of Federal Regulations and Document DO-260B, published by the Radio Technical Commission for Aeronautics, specify the position-integrity containment radius associated with NIC $\geq 7$ under normal operating conditions [47], [48]. Controlled receiver experiments reported in [49] further indicate that lower NIC values are consistent with GNSS degradation induced by interference.

Accordingly, a report is classified as degraded when NIC $\leq 6$. To reduce isolated aircraft-specific effects, a second is classified as ADS-B-positive only when at least $n$ aircraft within the WSMR analysis region simultaneously report NIC $\leq 6$, consistent with the concurrent-degradation approach described in [50]. As $n$ increased from 1 to 4, the proportion of ADS-B-positive seconds outside the NOTAM windows decreased from 48.90% to 19.30%, 7.46%, and 2.42%, respectively, while the separation between the in-window and out-of-window periods became progressively clearer. This pattern supports a temporal association between concurrent multi-aircraft degradation and the NOTAM schedule. We selected $n = 4$ because it produced the lowest out-of-window positive rate among these values, at 2.42%, while retaining a sufficient number of reference-positive seconds for quantitative evaluation. Increasing the criterion to $n = 5$ reduced the out-of-window positive rate by only about 2 percentage points, but decreased the number of reference-positive seconds by approximately 36%, substantially limiting the available sample size for quantitative analysis. This selection used only the NOTAM and ADS-B data, not the CYGNSS detector outputs. The four-aircraft condition was feasible at WSMR because of its dense ADS-B coverage; Section IV-D gives the corresponding traffic statistics for the Middle East.

A reference-positive second is defined by the intersection of the ADS-B-positive condition and the documented WSMR NOTAM windows. Seconds outside the NOTAM windows are treated as reference-negative, whereas in-window seconds without ADS-B corroboration are treated as indeterminate and excluded from the evaluation. Of the 40,841 s falling within the NOTAM windows, 6,148 s satisfied the ADS-B condition and were retained as reference-positive, while the remaining 34,693 s were excluded on this basis; the 93,098 s outside the windows constitute the reference-negative set. The evaluation therefore covers N = 99,246 s, of which 6.2% are reference-positive, and every reported PD is computed over the 6,148 reference-positive seconds while every reported FAR is computed over the 93,098 reference-negative seconds. Detector agreement with this reference standard is quantified using the probability of detection (PD), false-alarm rate (FAR), precision, and F1-score. At the daily scale, the mean CYGNSS DDM noise floor was strongly correlated with the proportion of ADS-B reference-positive seconds calculated over the same NOTAM-active intervals (Pearson $r = 0.895$, $p < 0.0001$). This association supports the interpretation that the two physically different observation systems capture a common interference-related temporal pattern.

Table II reports PD and FAR over thresholds from 40.0 to 41.5 dB. The 41 dB value selected in Section IV-A from the reference-region distributions is also the lowest tested threshold at which the proposed detector reaches an observed FAR of 0.0000, at a PD of 0.595; raising it to 41.5 dB holds the FAR at 0.0000 but lowers the PD to 0.553. At 41 dB the raw maximum raises PD from 0.532 to 0.631 relative to the mean baseline, while the kurtosis flag is sensitive but flags 72.71% of the reference-negative seconds. Precision and F1 further characterize this trade-off: the mean baseline and the proposed detector produce no false positives among the reference-negative seconds and attain a precision of 1.000, with F1-scores of 0.694 and 0.746, respectively. The raw maximum attains a precision of 0.997 and an F1-score of 0.773, whereas the kurtosis flag attains 0.070 and 0.128.

Screening costs little sensitivity. At 41 dB it retains about 94% of the raw-maximum PD, reducing it from 0.631 to 0.595, and lowers the observed FAR from 0.0001 to 0.0000 across the 93,098 reference-negative seconds. Persistence provides most of that suppression, reducing the raw-maximum FAR at every tested threshold. Concurrence alone reaches a PD of only 0.156 to 0.197, because simultaneous CYGNSS coverage of WSMR is infrequent, but its observed FAR is 0.0000 at every threshold, so it contributes a small number of independently supported detections without increasing the observed FAR. At approximately matched FARs the proposed detector retains slightly higher PD despite a 0.5 dB higher threshold: 0.647 versus 0.645, 0.595 versus 0.586, and 0.553 versus 0.532.

McNemar's test on the paired outcomes at 41 dB gives $\chi^2 = 321.3$ ($p < 0.001$), with 422 s detected only by the proposed detector against 37 s detected only by the baseline. A block bootstrap with 300 s blocks, which preserves temporal dependence among consecutive observations, gives a PD difference of +0.063 with a 95% confidence interval of [+0.034, +0.094]

TABLE II

DETECTION PERFORMANCE AND THRESHOLD-DEPENDENT ABLATION OF THE COMPARED METHODS AT WSMR, EVALUATED AGAINST THE NOTAM ∩ ADS-B REFERENCE STANDARD (PD / FAR; N = 99,246: 6,148 POSITIVE, 93,098 NEGATIVE)

| **Method** | **40.0 dB** | **40.5 dB** | **41.0 dB** | **41.5 dB** |
|---|---|---|---|---|
| Kurtosis flag (fixed, k > 4.0) | (0.798 / 0.727) | | | |
| Mean baseline | 0.645 / 0.0006 | 0.586 / 0.0000 | 0.532 / 0.0000 | 0.482 / 0.0000 |
| Raw maximum | 0.707 / 0.0098 | 0.675 / 0.0014 | 0.631 / 0.0001 | 0.590 / 0.0001 |
| Persistence only | 0.684 / 0.0078 | 0.645 / 0.0006 | 0.592 / 0.0000 | 0.549 / 0.0000 |
| Concurrence only | 0.197 / 0.0000 | 0.187 / 0.0000 | 0.170 / 0.0000 | 0.156 / 0.0000 |
| **Proposed (Max + screening)** | **0.686 / 0.0078** | **0.647 / 0.0006** | **0.595 / 0.0000** | **0.553 / 0.0000** |

PD: probability of detection; FAR: false-alarm rate.

### *D. Consistency in a Persistent-Interference Environment*

An equivalent epoch-level reference standard could not be constructed for the Middle East. Public interference schedules were unavailable, and ADS-B traffic was far sparser than at WSMR: over comparable numbers of epochs the WSMR region accumulated 994,039,366 aircraft-seconds against 31,490,811 in the Middle East, a mean of 479.4 against 15.2 aircraft per epoch. ADS-B coverage was therefore insufficient to construct a reliable reference-positive set for epoch-level quantitative evaluation. The Middle East analysis instead uses CYGNSS flagging proportions, over the same 1–24 May 2025 period, to examine whether the behavior observed at WSMR also appears in a second region widely reported to experience GNSS interference [7], [10].

The proposed detector produced the highest daily flag counts throughout the period (Fig. 8) and flagged 212,534 of 343,130 observed epochs (62%), against 156,487 (46%) for the mean baseline and 113,950 (33%) for the kurtosis flag (Table III).

TABLE III

RFI FLAG STATISTICS BY DETECTION METHOD OVER THE MIDDLE EAST (MAY 1–24, 2025)

| Detection Method | Flagged Epochs | Percentage of Total Epochs |
|---|---|---|
| Kurtosis flag | 113,950 | 33% |
| Mean baseline | 156,487 | 46% |
| Proposed | 212,534 | 62% |
| Total Epochs | 343,130 | 100% |

Fig. 9(a) shows Middle East events with the vertically enhanced pattern, and Fig. 9(b) shows cases with elevated background power and no such structure. Both share the feature identified in Section II: power elevated outside the nominal scattering region, including the forbidden zone. The detector therefore responds to both canonical vertical-stripe patterns and atypical elevated-background DDMs, rather than to a single visual morphology.

WSMR provides reference-based detector performance, whereas the Middle East shows that the same ordering of flagging proportions persists in an independently reported interference region for which no epoch-level reference exists

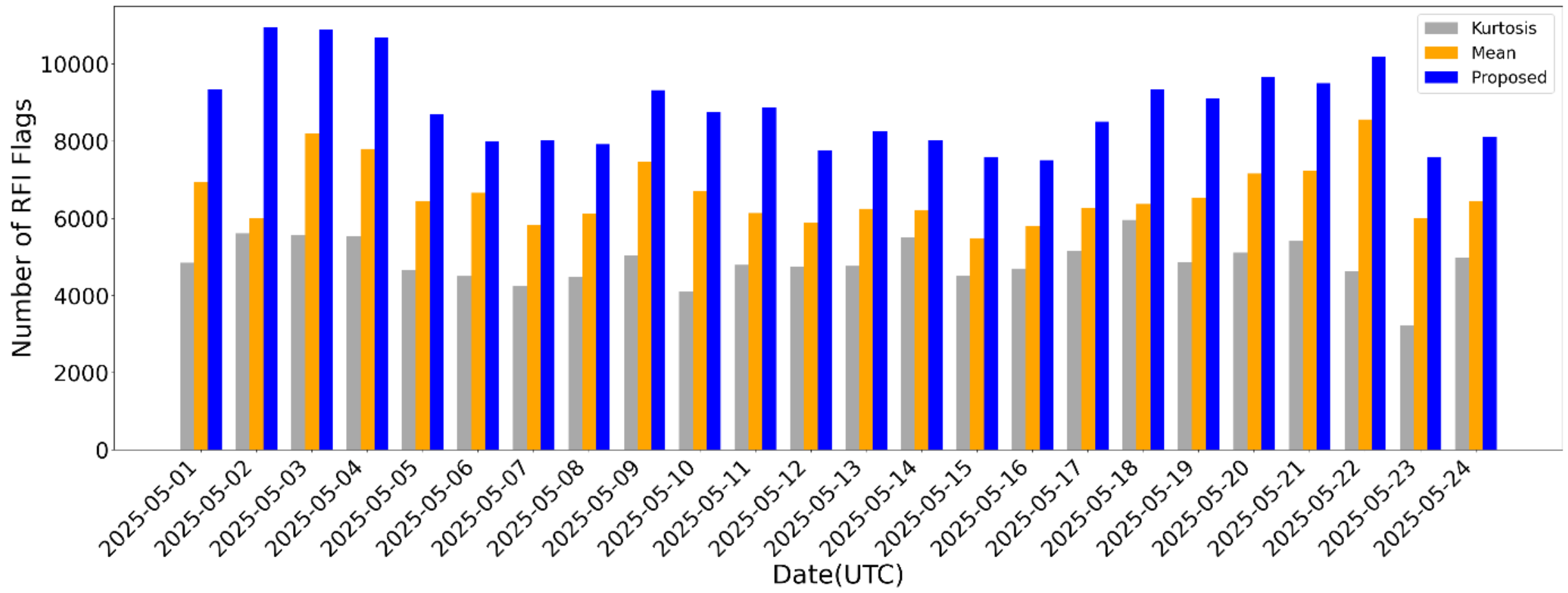

**Fig. 8.** Daily RFI flag counts over the Middle East from May 1 to 24, 2025, for the kurtosis flag, the mean baseline, and the proposed detector.

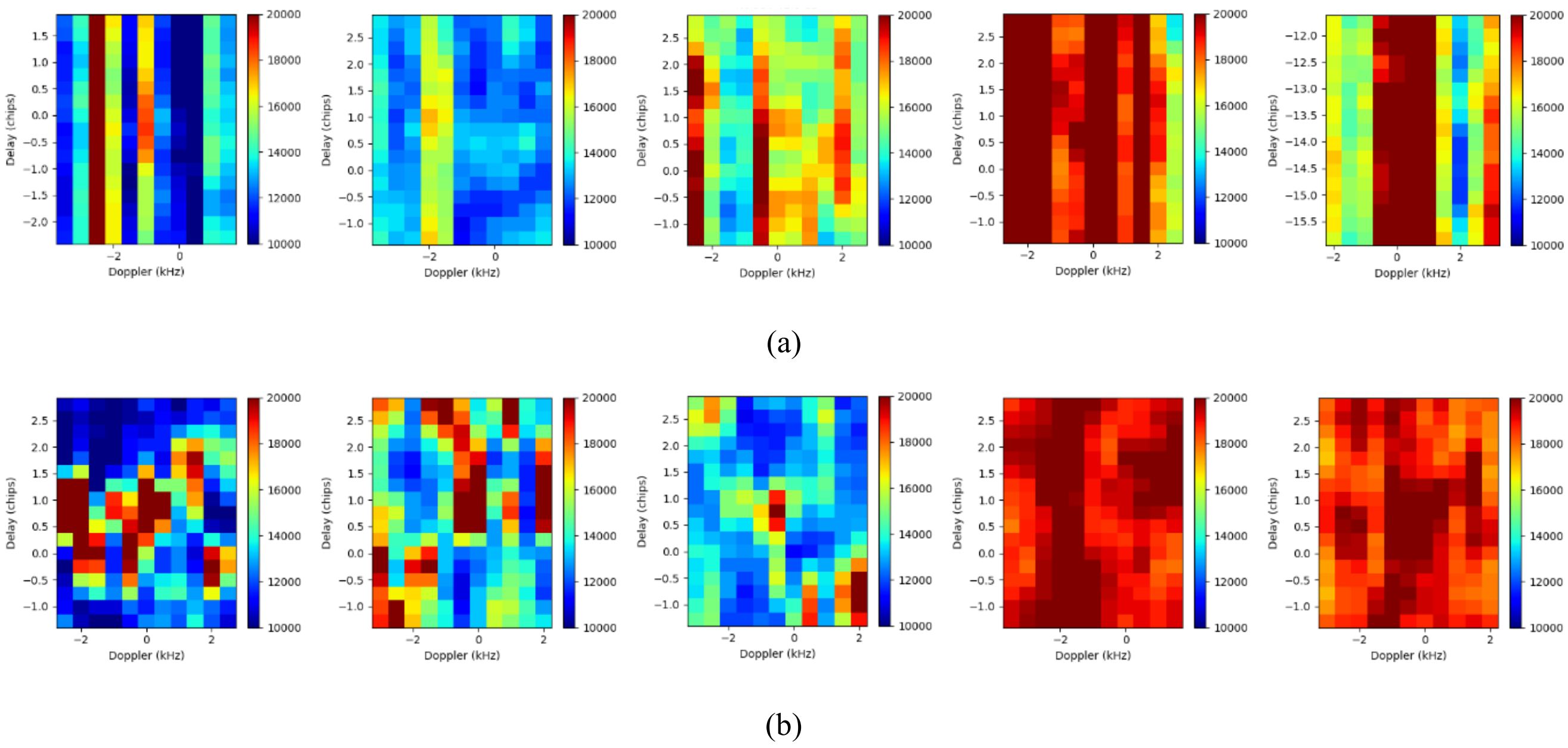

**Fig. 9.** Representative Middle East DDMs detected by the proposed detector: (a) vertical-stripe and (b) atypical elevated-background patterns.

## V. DISCUSSION

### *A. Channel Asymmetry and Detection Sensitivity*

Single-epoch performance depends directly on the channel-combination rule. The proportion of epochs with at least a 1 dB difference between the mean noise-floor levels of the two nadir-antenna groups rose from 14.9% below 40 dB to 75.5% at or above 40 dB, so the asymmetry that the maximum is designed to preserve is particularly common under the elevated-noise-floor conditions of interest. When the two antenna groups respond unequally, a source favorably placed with respect to one antenna and viewing direction is less likely to be diluted below threshold by the remaining channels. These measurements do not establish a one-to-one relationship between antenna-group asymmetry and individual detections, but they explain why the maximum is more effective than averaging under the conditions examined here.

### *B. Practical Applicability and Implications for Wide-Area Monitoring*

The detector runs on standard Level-1 products: four scalar values per epoch, one threshold comparison, and the subsequent logical screening. The per-epoch computation is negligible relative to Level-1 product delivery, making continuous processing of delivered products practical without raw-IF access or image-level classification. Detections included both canonical vertical-stripe patterns and atypical elevated-background structures, so the method is not tied to a single DDM morphology. The association between the CYGNSS noise-floor response and ADS-B navigation-integrity degradation, and the multi-satellite responses observed at WSMR, are natural starting points for future work on spaceborne–aviation data fusion and on constellation-level modeling of source observability.

The Middle East case illustrates the value of spaceborne monitoring where independent ground-based reference data are unavailable. Public schedules were unavailable there and ADS-B coverage is too sparse for an aircraft-based reference standard, yet the detector still produced a coherent daily record of CYGNSS noise-floor flags from Level-1 products. Spaceborne monitoring of this kind therefore provides wide-area situational awareness in regions where public schedules and dense terrestrial observations are unavailable.

### *C. Limitations*

The NOTAM–ADS-B reference is not an exhaustive epoch-level truth set. A NOTAM is a scheduling notice: it does not confirm that a transmitter was radiating during any particular second. The ADS-B condition narrows those windows to intervals in which aircraft reported degraded navigation integrity, but a NIC value reflects one aircraft's own position solution and depends on its avionics and flight geometry, and integrity can degrade for reasons unrelated to terrestrial interference. Requiring four aircraft to degrade concurrently reduces, but does not completely eliminate that influence. ADS-B and CYGNSS also observe different signals and geometries, so an ADS-B-positive second does not guarantee a measurable CYGNSS response, and the absence of corroboration within a window does not establish the absence of interference; such seconds are treated as indeterminate rather than reference-negative. The reported metrics are therefore measures of agreement between two structurally different observation systems, suitable for comparing detectors against a common reference but not for estimating absolute detection performance.

The 41 dB operating threshold was established from the reference-region distributions and subsequently evaluated through the threshold sweep at WSMR, where it was the lowest tested threshold reaching an observed FAR of 0.0000. Whether the same value carries over to regions with different terrain complexity, observation geometry, or receiver behavior was not tested here and would benefit from further validation. DDM-noise-floor elevation may also arise from delay-reference mismatch, topographic complexity, and other scene-dependent background effects. Some candidates rejected by the persistence criterion exhibited isolated one-to-three-epoch excursions followed by an immediate return to the surrounding background, with DDM morphology closer to nominal than to the representative contaminated patterns; their origin cannot be resolved from the available Level-1 observations. Adaptive thresholding based on local background statistics is a natural extension of the present approach and merits further study.

The current formulation is mission-specific. The analysis was performed using CYGNSS Level-1 products, and the threshold is defined on a logarithmic scale derived from CYGNSS instrument counts. Although the logic of taking the channel maximum is not inherently limited to CYGNSS, the numerical threshold and detailed response characteristics cannot be transferred directly to other GNSS-R missions without additional calibration and validation.

The 10 s persistence interval is a heuristic screening parameter. It corresponds to roughly 70 km of along-track motion at the CYGNSS ground-track speed of about 7 km/s, over which the slant range to a fixed ground source changes by only a few kilometres. The observable duration of a source also depends on its power, antenna pattern, and viewing geometry, none of which the framework normalizes, since it assumes no prior knowledge of jammer location. Sensitivity to this parameter should be evaluated in future work.

The results therefore support the maximum under the CYGNSS conditions evaluated here, rather than universal optimality across missions and environments.

## VI. CONCLUSION

This paper addressed how the four simultaneous CYGNSS Level-1 noise-floor channels should be combined to form a single-epoch decision variable, and how the resulting detector performs against an independent reference. Because the four channels are received through two differently oriented nadir antennas, elevated-noise-floor responses are frequently asymmetric across them; taking the channel maximum preserves the strongest response, and persistence and multi-satellite concurrence screening restrict retained events to wide-area or temporally sustained interference.

Against a reference standard combining the WSMR NOTAM schedule with concurrent ADS-B navigation-integrity degradation, the detector achieved a PD of 0.595 against 0.532 for the mean baseline, both at an observed FAR of 0.0000, and the difference was statistically significant. Screening retained about 94% of the raw-maximum PD while reducing its observed FAR from 0.0001 to 0.0000. In the Middle East, where no epoch-level reference exists, the same ordering of flagging proportions held. These results show that epoch-level wide-area interference monitoring is feasible from routinely distributed Level-1 products. Future work should address adaptive background estimation, observation-specific gain and geometry correction, sensitivity to the persistence interval, and validation across additional regions and GNSS-R missions.

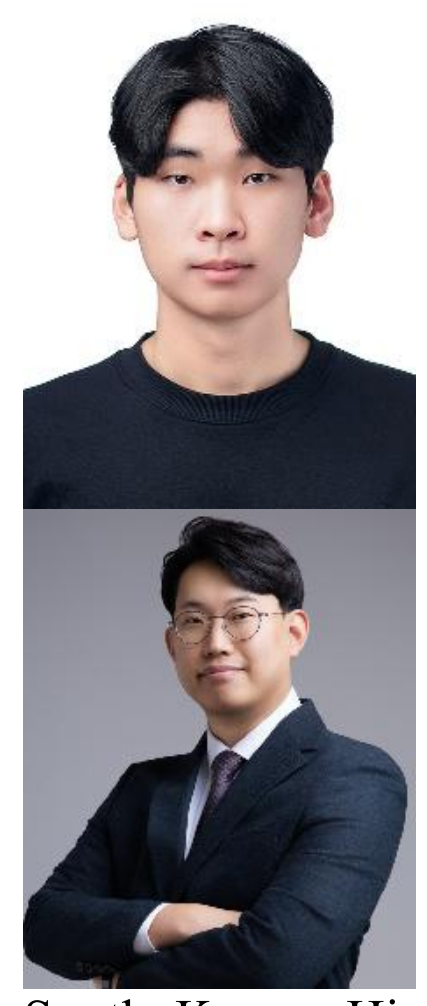

**Ji-Hyeon Shin** is currently working toward the B.S. degree in electronics engineering at Chungbuk National University, Cheongju, South Korea. His research interests include GNSS interference monitoring using Low Earth Orbit (LEO) satellites, satellite signal processing, GNSS-R, and GNSS integrity.

**Pyo-Woong Son** received his B.S. degree in electrical and electronic engineering from Yonsei University, Seoul, South Korea, in 2012, and his Ph.D. in integrated technology from Yonsei University, Incheon, South Korea, in 2019. He is currently an Associate Professor at Chungbuk National University, Cheongju, South Korea. His research interests include navigation and surveillance systems supporting resilient positioning, navigation, and timing (PNT) services.